# The AutoMat CVIM - A Scalable Data Model for Automotive Big Data Marketplaces

Johannes Pillmann, Benjamin Sliwa and Christian Wietfeld
TU Dortmund University, Communication Networks Institute (CNI)
Otto-Hahn-Str. 6, 44227 Dortmund, Germany
{johannes.pillmann, benjamin.sliwa, christian.wietfeld} @tu-dortmund.de

*Abstract*—In the past years, connectivity has been introduced in automotive production series, enabling vehicles as highly mobile Internet of Things (IoT) sensors and participants. The Horizon 2020 research project *AutoMat* addressed this scenario by building a vehicle big data marketplace in order to leverage and exploit crowd-sourced sensor data, a so far unexcavated treasure. As part of this project the Common Vehicle Information Model (CVIM) as harmonized data model has been developed. The CVIM allows a common understanding and generic representation, brand-independent throughout the whole data value and processing chain. The demonstrator consists of CVIM vehicle sensor data, which runs through the different components of the whole *AutoMat* vehicle big data processing pipeline. Finally, at the example of a traffic measurement service the data of a whole vehicle fleet is aggregated and evaluated.

## I. INTRODUCTION

With the growing integration of connectivity and data services into vehicles, cars are becoming an increasingly important part of the IoT [1]. Vehicles that drive around nearly in every corner of the world create a large-scale wireless sensor network and provide an enormous big data potential. In order to exploit this so far unexcavated treasure and provide an unhindered and unrestricted access-platform, the Horizon 2020 project AutoMat[1] created a vehicle big data marketplace (Fig. 1). On this marketplace, different car-manufacturers can participate. It provides open and standardized interfaces for data access and offers. Data is stored inside private data storage vaults (Fig. 3). Vehicle users need to give their consent in order to allow the marketplace brokering the data.

The AutoMat ecosystem enables new types of crowd-sourced services. One possible application is the road roughness detection using the sensors of millions of cars in Europe in order to detect potholes and improve the quality of streets. Within the scope of the project also non-automotive applications are evaluated by integrating vehicles into weather models. Especially in areas, with a low density of professional weather stations, where traditional models are not accurate, a significant impact can be achieved [2].

## II. COMMON VEHICLE INFORMATION MODEL (CVIM)

One key success factor of a data sharing ecosystem is a unified as well as efficient data representation. In order to harmonize proprietary car-manufacturer formats, the CVIM has been developed and evaluated as part of the AutoMat project [2], [3]. The data model is designed in a layered approach as shown in Fig. 2. On the bottom layer are the raw information providers within vehicles, e.g. an engine speed signal captured from the internal Controller Area Network (CAN) bus. As storage and data uplink bandwidth are limited (less technically, but more in terms of cost) a preprocessing of the raw in-vehicle data is necessary, which is performed by the inner *Measurement Channel Layer*.

[1]http://www.automat-project.eu

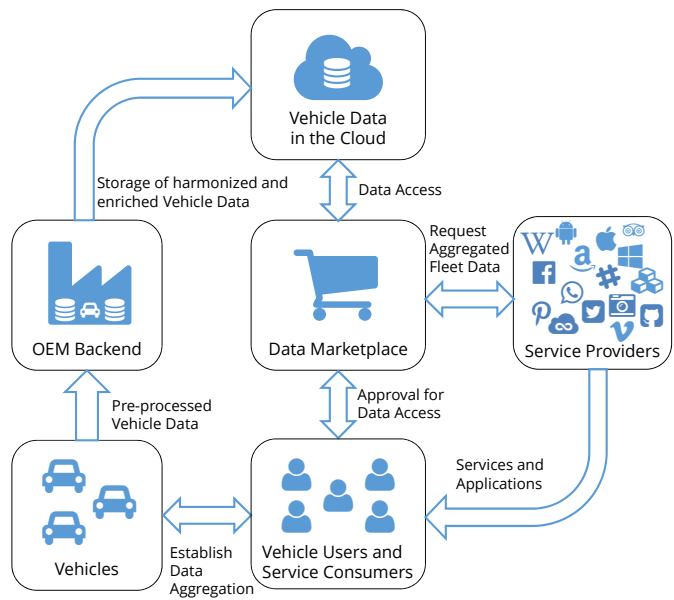

Fig. 1. AutoMat vehicle big data marketplace overview

Measurement Channels define an aggregation method of signals. CVIM supports time-series and histogram aggregation methods. For example, one measurement channel down-samples the $10\ ms$ CAN-bus engine speed signal to a $1\ s$ time-series, or stores its value distribution in a histogram via an additional measurement channel. Histograms have the advantage of requiring less data volume in comparison to time-series. On the downside, the time-dimensions in histograms are dropped and therefore a precise position, where the measurement was made, is lost. However, this aspect is still advantageous from the privacy perspective of data gathering as it reduces traceability of users and increases anonymity.

In order to still enable spatial correlation of measurements, while preserving the advantage of low data sizes, CVIM introduces the novel concept of *geo-histograms*. Here, the global map is segmented into a grid tiles. For each tile of the grid one histogram is created, whenever a car moves inside of the according cell. Those geo-histograms are represented using a sparse matrix representation.

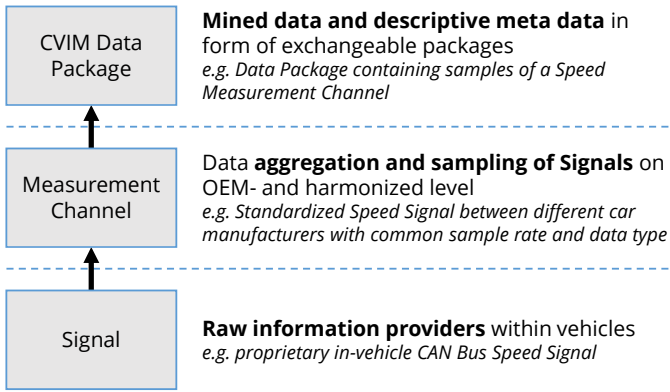

Fig. 2. Layered CVIM architecture

On the top layer are the *CVIM Data Packages*. They contain the actual aggregated vehicle sensor data. Next to the harvested time-series or histograms, they contain descriptive meta data. The metadata consists of descriptive properties, providing a rough estimate of the data recording time (start- and stop-timestamps) and place of recording (geo bounding boxes). The ownership information attribute describes different stake- or copyright holders of the stored data, whereas the car-manufacturer's checksum, sequence numbers and signature ensure the completeness and validity of data. The metadata supports the marketplace in indexing and structuring in order to allow an improved brokerage of data.

## III. USE-CASE & DEMO

The demonstration at the conference will present a showcase application leveraging a prototype implementation of the discussed CVIM data model. Conference attendees will be able to retrace the journey of CVIM data packages from their creation until the final aggregation and processing.

In the first part of the demo will address the creation of vehicle data. The open-source simulator *Lightweight ICT-centric Mobility Simulation (LIMoSIM)* [4] is leveraged to produce large scale vehicle big data of a $1000 \; cars$ fleet in the according CVIM format. All data will passed through the AutoMat ecosystem and marketplace broker and stored in the private cloud storage system.

The second part of the demo focuses on the processing and analytics of vehicle big data. The showcase service evaluates and aggregates CVIM data in order to show traffic bottlenecks and congestions. Hereby, the different aggregation types of trajectory-based time-series (video[2]) and geo-histograms (Fig. 4) are leveraged and explained in detail.

## ACKNOWLEDGMENT

This work has been conducted within the AutoMat (Automotive Big Data Marketplace for Innovative Cross-sectorial Vehicle Data Services) project, which received funding from the European Union's

[2]AutoMat SDK: service video, https://youtu.be/PddfoL3qToM

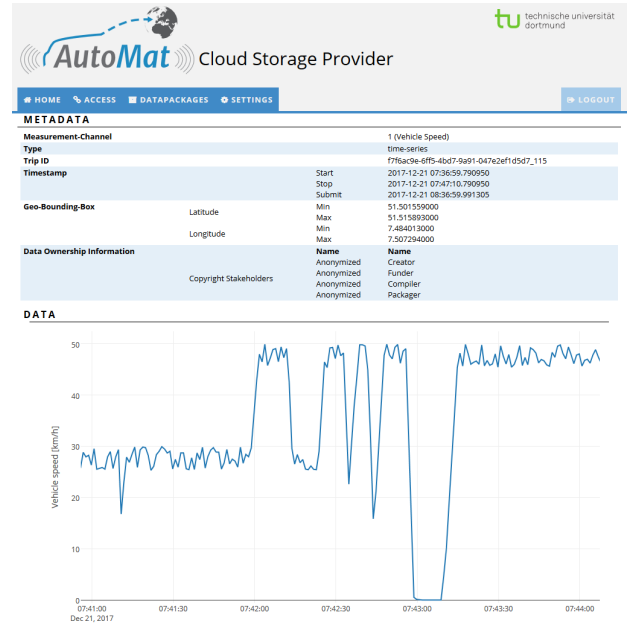

Fig. 3. Example CVIM data package inside its owner's private cloud storage vault showing a vehicle speed time-series.

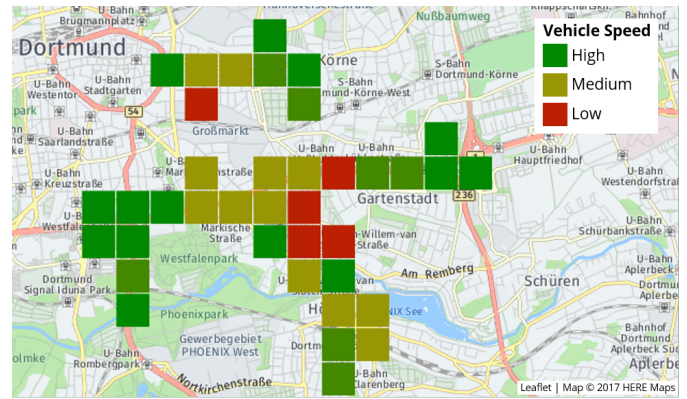

Fig. 4. Screenshot of geo-histogram based data aggregation. Green boxes indicate good traffic flow and red boxes congestion.

Horizon 2020 (H2020) research and innovation programme under the Grant Agreement no 644657, and has been supported by Deutsche Forschungsgemeinschaft (DFG) within the Collaborative Research Center SFB 876 Providing Information by Resource-Constrained Analysis, project B4 and European Regional Development Fund (EFRE) 2014-2020 in the course of the InVerSiV project under grant number EFRE-0800422.